# Evaluating Tooling and Methodology when Analysing Bitcoin Mixing Services After Forensic Seizure

Edward Henry Young, Christos Chrysoulas, Nikolaos Pitropakis, Pavlos Papadopoulos and William J Buchanan
School of Computing
Edinburgh Napier University
Edinburgh, United Kingdom
{40411848@live.napier.ac.uk}, {c.chrysoulas, n. pitropakis, p.papadopoulos, b.buchanan}@napier.ac.uk

*Abstract*—Little or no research has been directed to analysis and researching forensic analysis of the Bitcoin mixing or 'tumbling' service themselves. This work is intended to examine effective tooling and methodology for recovering forensic artifacts from two privacy focused mixing services namely Obscuro which uses the secure enclave on intel chips to provide enhanced confidentiality and Wasabi wallet which uses CoinJoin to mix and obfuscate crypto currencies. These wallets were set up on VMs and then several forensic tools used to examine these VM images for relevant forensic artifacts. These forensic tools were able to recover a broad range of forensic artifacts and found both network forensics and logging files to be a useful source of artifacts to deanonymize these mixing services.

*Keywords—Forensic Analysis, Bitcoin, Wasabi, Obscuro*

## I. INTRODUCTION

In 2021, US officials arrested and seized the servers of Bitcoin Fog who they accused of laundering over 1.2 Million Bitcoin which in turn had laundered over $330 million in real terms. When the FBI seized these servers, they will have been confronted by a complicated privacy-focused operation looking to obfuscate users' fund. Having been confronted by this bitcoin mixer and a system designed to obscure the source of the bitcoin and the identities of the users, law enforcement would have needed to utilise a different skill set and set of tools to trace the funds, completely unlike the forensic accounting techniques used for more traditional forms of money laundering.

Bitcoin was conceived as a peer-to-peer currency [1], where users can send and receive bitcoin via peer exchange. Users create a key pair, which consists of a public key which identifies the account to the world, and a private key, which is used to cryptographically sign transactions. Transactions list inputs and outputs. Inputs contain previous transactions which contain amounts of bitcoin. This is to evidence that coins have not already been transferred or spent.

Bitcoin uses a proof-of-work (PoW) system to verify transactions and to prevent double-spending. Conflicts in the system are resolved by majority decisions, with the weight of the vote based on computational power. Records of this spend are appended to the public record on the blockchain. To check that bitcoins have not already been spent, blockchain users keep an index of unspent transactions and reject those with invalid inputs from being integrated into a block. Bitcoin and other crypto currencies are generally thought to be anonymous and while the holder of wallets and coins are generally anonymous, the intricately public nature of the blockchain means that all transactions are stored publicly and are, by definition, available for inspection. Establishing the owner of the wallet holding the bitcoin or benefiting from its use can de-anonymise it. Mixing services are a reaction to this risk and aim to break the link between holder of the wallet and the origin of the coins by mixing the coins of multiple users, making it harder to find a relationship between input and output transactions. Bitcoin tumbling (or mixing) is used to provide an extra level of anonymity in bitcoin transactions. They work by mixing potentially identifiable or 'tainted' cryptocurrency funds with others, to obscure the trail back to the fund's original source. This addresses the privacy concern with bitcoin transactions that while the possessor or 'owner' of the currency might be anonymous, the transactions are by their nature recorded on the blockchain and visible for inspection. These services are characterised in [2] as providing anonymity via relationship anonymity.

This work will look to evaluate tooling and methodology when analysing bitcoin mixing services after forensic seizure and contribute to the current research base by examining what real world, publicly available tools and techniques uncover forensically as well as looking at the sources of artefacts that will require further academic attention. Its main focus will be in exploring: (a) What can be recovered from these services and what tooling is most effective? (b) How useful would this evidence be to investigators? (c) What additional techniques or approaches might generate best evidence for law enforcement? and (d) What mixing service techniques make them robust to forensic recovery?

## II. LITERATURE REVIEW

Current research in this area has focused on forensic analysis of the blockchain itself in general [3] or to identify mixing services via open-source intelligence from the blockchain [4].

It is possible to use Bitcoin addresses to identify users. The authors in [5] analysed structural aspects of the blockchain transaction to graph and draw implications on the anonymity of transactions. Similarly, the authors in [6] studied privacy implications of multi-input transactions and shadow addresses generated by the Bitcoin client for receiving change. The method by which the block chain is analysed to determine which bitcoin addresses are related to others is called taint analysis.

Forensic blockchain analysis has paid dividends. Meiklejohn et al. [7] could identify 40% of the users in an artificial transaction graph based on (simulated) behaviour. It was possible to identify many intermediaries by interacting with them and using a change address heuristic to discover addresses of the same user. Using this dataset, they analysed popular thefts and related pay-outs to popular exchanges. Seizing these servers and recovering information appropriately will help unlock a significant amount of anonymity to these transactions. If investigators have knowledge of bitcoin addresses owned by both the person of interest and the third-party mixing service, they can identify the transactions between the two. Users must trust that the service has enough customers in order to effectively mix the number of bitcoins they have deposited and that they do not retain any log files of the mixing. The above indicates that at least in some cases, the mixing services may not be as secure as they suggest to their users, with transactions of services being able to be picked out of the blockchain or services being hacked with some degree of ease.

Wasabi wallet is one of the officially recommended desktop Bitcoin wallets and one of the few with integrated CoinJoin[1] functionality. Wasabi wallets are increasingly important in law enforcement investigations. The EU 2020 assessment of cybercrime noted that privacy-enhanced wallet services using CoinJoin concepts such as Wasabi have emerged in their assessment as a top threat in addition to well established centralised mixers (IOCTA 2020). Wasabi wallet also conducts all communications via TOR providing additional anonymity to transactions.

CoinJoin is a response by the bitcoin core developers to provide greater relationship anonymity to bitcoin users. This service does not rely on a centralized mixing platform. ConJoin arranges structured transactions merging different inputs and outputs in a single transaction. This makes matching transactions much more difficult to track which input 'pays' which output, thus making attribution difficult. Outputs in a CoinJoin must be equal to ensure that all users receive the same value output when the process completes. If there are differences in pay-outs, this could lead to deanonymisation. Crucially for bitcoin forensics, the coordinator of a CoinJoin has insight into users' information that could allow them to link inputs to a user. This opens up the potential of uncovering important artefacts if a Wasabi Wallet is forensically analysed. This role will carry on even when Wasabi Wallet[2] 2.0 is put into production.

Obscuro takes a similar approach, using a de-centralised mixing service, albeit one which employs distinct anti-forensic techniques. Obscuro employs hardware-based trusted execution environments (TEE) to protect its mixing operations from its operating environment. The content of the secure enclave is encrypted and stored in RAM, leaving theoretically minimal artefacts on the drive of the system in which it operates, providing strong memory integrity and confidentiality. Obscuro utilises Intel's SGX system secure enclave to separate application data from the rest of the system. Grundmann et al [8] have called this and similar services of TEEs BANKLAVES. Examples include Tesserect, Teechain and Obscuro.

Banklaves generally work by creating a new network of secure enclaved mixers. Typically, they work as follows: Alice starts her client creating the first enclave Ea. Bob joins the network by letting his enclave Eb perform a remote attestation with Ea. To deposit, Alice's enclave to creates a deposit address. Then, Alice creates a Bitcoin transaction to this address, publishes to the blockchain and gives the transaction to her enclave. Alice checks that the transaction is part of a blockchain and increments Alice's balance in the state. For transfers between Alice and Bob, Alice calls the function in her enclave and passes the amount and the receiver of the transaction. Ea verifies that Alice has enough balance available for that transaction and then updates the balances and sends them to Eb. If Alice wants to cash out, her enclave creates a Bitcoin transaction that she can publish on the blockchain. This transaction spends cAlicebitcoins from the unspent transaction outputs managed by the enclave network to an address provided by Alice Ab [9].

The benefit of TEEs is that they are created with the assumption that in the event of a compromised OS, the enclave is secure and separated from the attacker, barring side channel attacks. The work in [8] outline as one of the key security objectives of TEEs that they aim to keep user balances and their performed transactions confidential from curious attackers. Bentov et al in [9] created Tesseract based on the assumption that an adversary (potentially the exchange operator) can gain complete physical access to the host in which the funds are stored and complete control of its network connections.

There is a surprising lack of research on the actual security of these systems. Most research has looked at various different models for implementation (e.g., Tesseract, Teechain, Obscuro) with differing attack vectors based on manipulation of inputs and outputs and unfaithful parties. There is little or no academic research currently available which looks at what forensic artefacts such mixers leave and how robust these wallets and mixing services are to modern forensic techniques. This paper is designed to fill in these research gaps.

III. METHODOLOGY

A. *Creation of mixers – Choice of VM, steps to Creation and Issues with Deprecation*

This project used virtualisation in order to facilitate the creation of these mixing services as well as making imaging and analysis simpler. This is also more reflective of the environment these services are likely to be run in, namely virtualised on cloud service providers. Two separate VMs were created to facilitate this experiment.

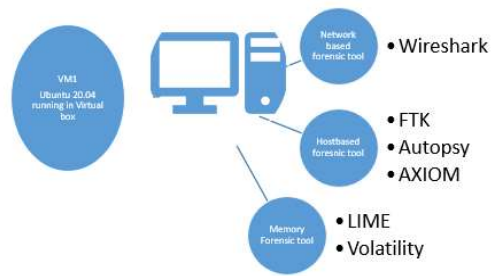

Fig. 1. VM1 to host Wasabi

VM1 was used to host Wasabi wallet. A virtual machine was created in virtual box (as this allowed for snap shotting of progress). This was compiled and built on Ubuntu 20.04 desktop and used an AMD Ryzen7 3.5 GHZ using 1 core and 8gb of memory Wasabi wallet was constructed in line with its instruction on Github (See Fig. 1).

VM2 was used to host Obscuro. This similarly was created as a VDX and mounted in virtual box again as this allowed snap shotting to take place. Obscuro had an issue with compilation as some of its components were being deprecated. A VDX was constructed with Obscuro and was built on an

---

[1] https://coinjoin.io/en

[2] https://wasabiwallet.io

Ubuntu 16.04.3 LTS Server 64bits operating system using 1GB of memory and 1 CPU core on an Intel I7. 1.8GHZ processor. Due to issues with deprecation of certain elements of Obscuro, this was the only operating system it was configured to work appropriately under. As this was a server version, a GUI was installed over the top in the form of a GNOME UI (See Fig.2).

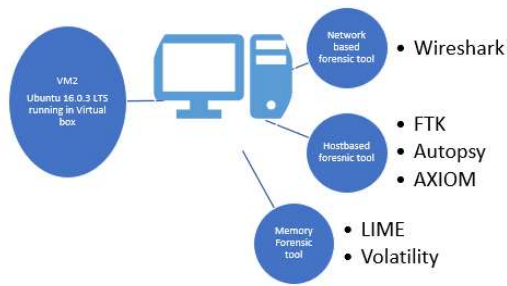

Fig. 2.  VM2 to host Obscuro

### B. Integration of Testnet/Regtest data and creating data for Forensic Recovery

Both VMs were started and the services of both Obscuro and Wasabi wallet started, had Regtest coins inserted into them and mixing services started. Regtest bitcoins are user created coins based on a private localised block chain with the same basic rules as bitcoin. As they are isolated there is no external communication, and this can be used as a quick and effective method for ascertaining how the mixing service works.

Mixing with Regtest coins proved successful so the experiment proceeded to use Testnet coins to mix within these services. Testnet is an alternative blockchain which can be used for testing. It operates like normal bitcoin, but the coins purposely hold no value. They need to be mined just like normal bitcoin and operate on a separate public blockchain. Due to the requirement for mining, Testnet coins are more difficult to come by but are available from several public faucets.

Testnet coins were used for this experiment as they provide a good approximation of true operation of these mixing services. Testnet mixing requires external communication with peers on the blockchain and operates identically to how real bitcoin and mixing would occur. They have the benefit of being free and available, designed for using during testing (so not linked to personal accounts of other users) and identical to Mainnet coins. Both VM's and services then began mixing using Regtest. Obscuro was cycled a number of times using Regtest coins (3 x 100) to generate sufficient test data to recover. Subsequently, Testnet coins were used imported to simulate network traffic.

Three Wasabi wallets were created and Testnet coins transferred into these. These wallets then CoinJoined between themselves and peers on the Testnet blockchain until the coins were mixed by this process.

### C. Choice of Forensic Tools – Suites Vs Standalone Tools

An investigation was made into existing tools to analyse VM1 and VM2. There are few to no standalone forensic tools designed to examine wallets or mixing services. There are a great multitude of tools such as chain analysis and blockchain explorer which are excellent at visualising and linking public blockchain activity and relationships and making this easy to use. As previously discussed in this work, there is little research conducted into analysis of wallets and mixing services themselves and consequently there are few standalone tools designed to provide forensic analysis or recovery of key artefacts.

Instead, Forensic suites would be used, and then specific forensic artefacts searched for within these tooling sets. This would have the benefit of allowing numerous tools to be used to examine for significant artefacts as well as providing advanced capabilities such as file and data stream carving, signature analysis and more enhanced parsing of specific applications.

Autopsy was chosen as one of the suites to be used. Autopsy is one of the main open-source digital forensics platform allowing analysis of hard drives, smart phones, media cards, etc. It is designed to offer analysis which is very similar to more premium, paid for suites in terms of capability while remaining open-source and, crucially, adaptable by virtue of its support for plugins which allow people to add functionality and, where appropriate, for these to be authorised by its maintainers as supported plugins. Crucially, Autopsy maintains a supported third party module for Bitcoin and wallet forensics.

FTK imager was the second suite used on VM1 and VM2. FTK Imager is an open-source software by AccessData which was chosen for its powerful ability to mount and parse image files. It is robust and able to deal with Linux/Unix OS's effectively as well as Windows operating systems.

Axiom is a paid-for, comprehensive suite of tools. Axiom allows images of devices to be captured, processes the images to recover data and provides analytical tools. It has significant capabilities to automate carving and parsing via machine learning as well as providing advanced capabilities such as mapping cross case correlations, and using advanced keyword searches, filtering and tags. Axiom has quickly come to dominate the field in terms of law enforcement forensic tooling and represents a more comprehensive tool than its niche but well-respected predecessor, IEF finder.

## IV.  IMPLEMENTATION AND RESULTS

With the bitcoin mixing services in operation and a test run on a multi-pronged approach to forensic recovery was then attempted. The design of this experiment was as follows. Each VM was analysed using a number of approaches using standard available tooling and examining for the below key artefact types: (a) Wallet data: For example Wallet.dat files that will provide the first transaction details, its current balance and the total amount received in that wallet since it was first seen; (b) Peer data: For example peers.dat files that contains IP addresses of peers the mixing service has connected to; (c) Public and private keys. This data will provide significant ownership evidence since the private key should only be available to the holder of the wallet; (d) Bitcoin transaction data. This is extremely important information for both matching transactions across the blockchain, but also for providing essentially logging data of what the mixing service has done to obfuscate the transactions; (e) Key word searching using the following regular expression which should highlight bitcoin artefacts [ For Bitcoin wallet addresses: ^[13][a-km-zA-HJ-NP-Z1-9]{25,34}$ and for Bitcoin private key strings: ^[5KL][1-9A-HJ-NP-Za-km-z]{50,51}$]

## A. Wasabi Wallet (VM1)

### 1) Autopsy

The .VDMK was loaded into Autopsy (4.17.0) and VM file selected as a data source. The VDMK was selected and auto detect selected for sector size. Autopsy comes with several built-in ingest models. All of these were selected including the Bitcoin FEA plugin which keyword searches for relevant bitcoin artifacts. The VDMK was then ingested and indexed by Autopsy. The FEA plugin produced no additional hits. Further keyword searching (wallet.dat,peer.dat,public key, private key, info.dat,bitcoin,coin) was expanded to look for significant artifacts. This yielded no additional worthwhile results though.

While keyword searching generated numerous hits, upon examination these were not relevant nor provided interesting forensic artefacts in relation to the operation of Wasabi Wallets. A file of note was recovered using the key word 'wallet.dat'. This file Name achilles_achilles.png appears to be a debugging file. Its unusual name might reflect some knowledge logging that in turn might reveal sensitive information that would undermine its confidentiality. Analysis of this file showed that it did contain some sensitive information (e.g., hash values and geneses block details), but did not reveal any sensitive information that could be used to undermine the anonymization of the mixing service or reveal information about wallet operation.

### 2) FTK

The .VDMK was loaded into FTK imager (4.17.0) and the VM file selected as a data source. FTK was unable to identify Wasabi wallet or any artefacts related to it. While it was able to identify the partition correctly, it could not identify anything underneath the parent home folder in which Wasabi was installed. It is highly likely some of this data has been wrongly assumed by FTK to be unallocated space. This could possibly be down to FTK having trouble interpreting the MFT of the VDMK either because of the variant of Ubuntu used or due to its virtualised nature. Regardless, it had a negative effect on forensic recovery.

### 3) AXIOM

The .VDMK was loaded into AXIOM (v3.5.1.1) which was running in the SIFT workstation which in turn was running in VM workstation 15. The Axiom process was used to process the VDMK and then loaded into Axiom Examine for further analysis. All evidence categories were selected to ensure that as much relevant information was processed for analysis as possible. Axiom contains several preconfigured artefact searches including for Bitcoin.

Axiom was not able to identify bitcoin transactions unlike with Obscuro. It was able to identify the use of TOR on this VM which is utilised by Wasabi wallet for traffic to peers and the blockchain. The absence of it being able to identify bitcoin transactions could possibly be explained since the wallet itself is encrypted as well as the private key with BIP: 38. As such, file or data stream carving would not be feasible as the contents would be obfuscated. See Table I.

TABLE I.  KEYWORD SEARCH RESULTS WITH AXIOM

| Keyword | Positive Hits | Relevant Material |
|---|---|---|
| Wallet.dat | No | No |
| Peer.dat | No | No |
| Public Key | No | No |
| Private Key | No | No |
| Info.dat | No | No |
| Bitcoin | Yes (3) | No |
| Coin | No | No |
| Bitcoin wallet addresses: ^[13][a-km-zA-HJ-NP-Z1-9]{25,34}$ | Yes (2) | No |
| Bitcoin private key strings: ^[5KL][1-9A-HJ-NP-Za-km-z]{50,51}$ | No | No |

## B. Obscuro (VM2)

### 1) Autopsy

The .VDMK was loaded into Autopsy (4.17.0) and VM file selected as a data source. The VDMK was selected and auto detect selected for sector size. Autopsy comes with several built-in ingest models. All of these were selected, including the Bitcoin FEA plugin which keyword searches for relevant bitcoin artefacts. The VDMK was then ingested and indexed by Autopsy. Autopsy appeared to have difficulty reading the image effectively as it seemed to believe the image to be almost entirely unallocated space. This is likely to be related to Obscuro's use of a server-based version of Ubuntu that Autopsy is not configured to process properly. However, Autopsy was able to parse the image and recover significant file structure including the Wasabi wallet parent folder. The FEA plugin produced no relevant hits as demonstrated below.

Further keyword searching (wallet.dat,peer.dat,public key, private key, info.dat,bitcoin,coin) was expanded to look for significant artefacts. This yielded some additional results as can be seen in Table II.

TABLE II.  KEYWORD SEARCH RESULTS WITH AUTOPSY

| Keyword | Positive Hits | Relevant Material |
|---|---|---|
| Wallet.dat | Yes (4) | Yes |
| Peer.dat | No | No |
| Public Key | Yes (711) | All false positive |
| Private Key | Yes (778) | All false positive |
| Info.dat | No | No |
| Bitcoin | Yes (116) | Yes (3) |
| Coin | Yes (47) | All false positive |

A key word search for "Wallet.dat" recovered 4 files of interest in unallocated space. Bitcoin contains log data but no information of relevance. The remaining 3 files $CarvedFiles/f1830504.db,Unalloc_509_512753664_10736369664 and /$carvedfiles/f5855600.txt contained identical fragments of x64 assembly code which appear to be in dump format which shows transactions, public keys and transaction hashes. A key word search for "Bitcoin" recovered 116 files. Of particular interest was a recovered database file CarvedFiles/f1830504.db. Examination of this db appears to be a chainstate db. This db will provide a representation of all currently unspent transaction outputs and some metadata about the transactions they are from. This is used to validate incoming blocks and transactions. A truncated example is produced below. This again is an example of the confidentiality of this environment being undermined.

*2) FTK*

The .VDMK was loaded into FTK imager (4.17.0) and the VM file selected as a data source. FTK could recover some forensic artefacts of interest and was able to read the OS and File system structure of the VM much more effectively and present a better overview of the file structure than Autopsy. Almost immediately, it was possible to pick out important files for forensic analysis of bitcoin miners. Namely, it was trivial to identify and recover peers.dat and wallet.dat which as previously discussed contain significant information about bitcoin held on the VM as well as peers on the network it had connected with. This wallet.dat file could be trivially imported into the bitcoinqt client or pwallet.py and significant forensic artefacts like transactions. Critically, the private key details could be extracted and be mounted by anyone looking to examine the files for further forensic evidence. An example of recoverable data is produced below. Wallet.dat was copied to a new Linux VM and pywallet.py used to extract relevant data. This trivially revealed sensitive data such as secret and private keys.

*3) AXIOM*

The .VDMK was loaded into AXIOM (v3.5.1.1) which was running in the SIFT workstation running in VM workstation 15. The Axiom process was used to process the VDMK and then loaded into Axiom Examine for further analysis. All evidence categories were selected to ensure that as much relevant information was processed for analysis as possible. Axiom contains a number of preconfigured artefact searches, including those relating to bitcoin transactions which it classifies as "Peer to Peer". AXIOM was able to recover a significant amount of additional general forensic artefacts versus Autopsy and FTK, including numerous emails and details of browser activity. In particular, it was able to successfully carve an excel file containing bitcoin debug logs in physical sectors 2832113 to 6857201. This carved document contains detailed bitcoin transaction logs, including wallets being loaded and transactions being sent and received. As previously detailed in this dissertation, this information undermines the confidentiality of the Obscuro executing transactions within its trusted execution environment when these are recoverable from debug logs the mixer is maintaining. Further keyword searching (wallet.dat,peer.dat,public key, private key, info.dat,bitcoin,coin) was expanded to look for significant artifacts. No useful findings returned though.

## V. CONCLUSIONS

We met our generalised aims in that we were able to report on significant artefacts recovered from the mixing services. Particularly, we were able to recover peers.dat and wallet,dat from Obscuro via FTK which are significant artefacts. We were also able to recover interesting logging data from both mixers which could be used to undermine the confidentiality of these services. Axiom was unsurprisingly (given its ubiquity within law enforcement and its cost as a closed source product) able to produce the most artefacts and many of the most relevant ones. The most effective anti-forensic techniques employed by the mixing services involved encrypting data at rest and in transit. While Obscuro mixed in the secure enclave its subsequent writing to disk of this data in encrypted artefacts made this on silicon mixing for security slightly redundant.

For Bitcoin users looking to ensure privacy and confidently for their transactions, this research provides a few important considerations. Users should look for mixing services that emphasise encryption at rest for their wallets and encryption in transit for their transactions. Doing so greatly limits the attack surface for forensic recovery of such transactions. Wasabi's emphasis on this makes it an especially strong product in this regard. Obscuro, while offering an effective and secure mixing process, neglects to adopt defence in depth for its confidentiality. While the use of TOR by Wasabi is generally an assistance in these privacy and security concerns, users should still be aware that threat actors can and do target TOR nodes looking for bitcoin traffic in attempts to steal funds or reveal users.

For those operating mixing services, the above holds equally true, but they should also look to proactively review their services and the wider operating system for what data created via any supporting frameworks their mixing services are logging, either directly or indirectly. That both Obscuro and Wasabi had log files which revealed information which could deanonymize transactions is indicative that this problem is widespread and could be used against multiple operators. This review would likely have to be done as part of penetration testing against these services on a regular basis to ensure that as the application, frameworks and the underlying operating systems are updated, new sources of logging are not created which could be recovered and used. In order to protect their users' confidentiality, mixers should proactively review their services for inadvertent log collection, encrypt wallets at rest and transmit data via TOR or some other form of encrypted communications channel.